\documentclass[superscriptaddress,aps,prd,preprintnumbers,nofootinbib,preprint,12pt]{revtex4}
\usepackage[utf8]{inputenc}
\usepackage[citecolor=blue,urlcolor=blue,unicode=true,pdfusetitle,
bookmarks=true,bookmarksnumbered=true,bookmarksopen=true,bookmarksopenlevel=3,
breaklinks=false,pdfborder={0 0 1},backref=false,colorlinks=true]{hyperref}
\usepackage{bm}
\usepackage{latexsym}
\usepackage{dcolumn,xcolor}
\usepackage[normalem]{ulem}
\usepackage{amsmath,amsfonts,amssymb}
\usepackage{graphicx,epsfig}
\usepackage{soul}
\usepackage{ulem}
\usepackage{amsthm}
\usepackage{color,float} 

\usepackage{amssymb}
\usepackage{braket}
\usepackage{physics}

\usepackage{tikz}
\newcommand{\orcidicon}{%
	\begin{tikzpicture}
	\draw[lime, fill=lime] (0,0) 
	circle [radius=0.16] 
	node[white] {{\fontfamily{qag}\selectfont \tiny ID}};
	\draw[white, fill=white] (-0.0625,0.095) 
	circle [radius=0.007];
	\end{tikzpicture}	\hspace{-2mm}
}

\newcommand\orcidSebastian{{\href{https://orcid.org/0000-0001-6235-120X}{\orcidicon}}}
\newcommand\orcidJames{{\href{https://orcid.org/0000-0002-3619-2505}{\orcidicon}}}

\begin{document}
	\title{ 
		Quantum Weak Equivalence Principle and the\\ Gravitational Casimir Effect in Superconductors }

	\author{Sebastian Bahamonde\orcidSebastian\!\!}
	\email{sbahamonde@ut.ee}
	\affiliation{\scriptsize{Laboratory of Theoretical Physics, Institute of Physics, University of Tartu, W. Ostwaldi 1, 50411 Tartu, Estonia.}}
	\affiliation{\scriptsize{Int.Lab. Theor. Cosmology,
			Tomsk State University of Control Systems and Radioelectronics (TUSUR), 634050 Tomsk, Russia}}
	\author{Mir Faizal}
	\email{mirfaizalmir@googlemail.com}
	\affiliation{\scriptsize{Department of Physics and Astronomy, University of Lethbridge, 4401 University Drive, Lethbridge, Alberta T1K 3M4, Canada}}
	\affiliation{\scriptsize{Irving K. Barber School of Arts and Sciences, University of British Columbia - Okanagan, 3333 University Way, Kelowna, British Columbia V1V 1V7, Canada}}
	\affiliation{\scriptsize{Canadian Quantum Research Center, 204-3002, 32 Ave,
			Vernon, BC, V1T 2L7, Canada}}
	\author{James Q. Quach\orcidJames\!\!}
	\email{quach.james@gmail.com}
	\affiliation{\scriptsize{Institute for Photonics and Advanced Sensing and School of Chemistry and Physics,
			The University of Adelaide, South Australia 5005, Australia}}
	\author{Richard A. Norte}
	\email{r.a.norte@tudelft.nl}
	\affiliation{\scriptsize{Department of Precision and Microsystems Engineering, Delft University of Technology, Mekelweg 2, 2628CD Delft, The Netherlands}}
	\affiliation{\scriptsize{Kavli Institute of Nanoscience, Delft University of Technology, Lorentzweg 1, 2628CJ Delft, The Netherlands}}
	\date{\today}
	\begin{abstract}
	We will use Fisher information to properly analyze the quantum weak equivalence principle. We argue that gravitational waves will be partially reflected by superconductors.
	This will occur as the violation of the weak equivalence principle in Cooper pairs is
	larger than the surrounding ionic lattice. Such reflections of virtual gravitational waves
	by superconductors can produce a gravitational Casimir effect, which may be detected
	using currently available technology
		\\
		{}
		\\
		{}\\
		{\textbf{Essay received an honorable mention for the Gravity Research Foundation 2020 Awards for Essays on Gravitation}}
	\end{abstract}
	\maketitle
\newpage
General relativity  is built upon the  weak equivalence principle (WEP), which has been demonstrated to be valid up to $\delta \sim 10^{-15}$ in precision (where $\delta$ in the E\"otv\"os parameter)~\cite{Touboul:2017grn}. The WEP states that due to the equivalence between the gravitational and inertial masses, the trajectory  of a classical particle in a gravitational field  does not depend on the mass of that particle.  Now for a quantum particle, trajectories are replaced by quantum probability distributions, which  in principle can violate  the WEP. However, 
the violation  of the WEP has not been observed in several quantum systems placed in constant  gravitational fields \cite{Lebed:2019myf,Lebed:2017itc,Zych:2015fka, Geiger:2018xwr,Rosi:2017ieh}.  
As these quantum systems  were studied  using constant gravitational fields, it is still possible that the WEP can be violated in varying gravitational backgrounds, like gravitational waves.
Thus, we will discuss the violation of the WEP  by quantum particles in a gravitational wave (GW), and  to do this we need to precisely define the quantum analog for the WEP.

The absence of information about the mass of the particle from its classical trajectory can be generalized to probability distributions to  obtain a definition for  quantum WEP. 
This is done using Fisher information, which measures the amount of information that an observable random variable provides about an unknown parameter. Now for a particle,  the random variable would be its measured position $\mathbf{x}$, and the unknown parameter would be its mass $m$. So, for  a quantum  particle  with wave function $\psi(\mathbf{x},t)$, the Fisher information is given by \cite{Seveso:2016qwh,Seveso:2016sct}
\begin{equation}
	F_x(m)=\int d\mathbf{x}|\psi(\mathbf{x},t)|^2
	\left[\frac{\partial}{\partial m}\log\left|\psi(\mathbf{x},t)\right|^2\right]^2~.
\label{eq:fisher}
\end{equation}
One notes that it is possible to obtain the mass information from the position of a quantum particle.  This mass information for a free particle $F^\text{free}_x(m)$,  represents the Fisher  information of the inertial mass (as the gravitational field is not present). However, for a particle in a  gravitational field, the mass information   $F_x(m)$, represents the Fisher information of the gravitational mass.
 So,  for the quantum WEP to hold, the Fisher information  in 
a gravitational field  $F_x(m)$ should be exactly equal to the Fisher information in the absence of such  a gravitation field
$F^\text{free}_x(m)$.  In fact, it can be observed that for constant  gravitational fields, $F_x(m)=F^\text{free}_x(m)$ \cite{quach2020weak}, and so the WEP holds, as expected from previous observations~\cite{Lebed:2019myf,Lebed:2017itc,Zych:2015fka,Geiger:2018xwr,Rosi:2017ieh}.

However,  we  can demonstrate that  new mass information can be obtained for a quantum particle in a GW. We start from  the metric for a  generally polarized linear plane GW 
\begin{equation}
ds^2=-c^2dt^2+dz^2+(1-2v)dx^2+(1+2v)dy^2-2udxdy~,
\label{gw_metric}
\end{equation}
where $u=u(t-z)$ and $v=v(t-z)$ are functions which describe a wave propagating in the $z$-direction. Now  for  a circularly polarized GW traveling along the $z$-direction, $v=f=f_0\cos(kz-\omega t)$ and $u=if$.  We can write the 
non-relativistic   Hamiltonian for a  quantum particle (using the Foldy-Wouthuysen transformation \cite{Obukhov:2000ih, Jentschura:2013dea} of the Dirac equation in curved spacetime) in the  GW background as \cite{Quach:2015iea,Quach:2016uxd}
\begin{eqnarray}
H_\text{GW}=\frac{1}{2m}(\delta^{ij}+2T^{ij})p_ip_j + mc^2 ~,
\label{eq:H}\quad
T_i^j=
\begin{pmatrix}
v& &-u &0\\
-u& &-v &0\\
0& &0 &0\\
\end{pmatrix}~.
\label{eq:T}
\end{eqnarray}
The unitary transformation
$U=\exp[-i/\hbar\, \int H_\text{GW}(t)dt]$ gives the time evolution 
of 
 a localized quantum particle (with a one dimensional wave packet   $
	\psi(x,0) = 
	({2}/{\pi})^{1/4}e^{-(x-x_0)^2}~
$)  
in a GW background  \cite{Quach:2015iea, quach2020weak}
\begin{eqnarray}
	\psi(x,t)=\Big(\frac{2}{\pi}\Big)^{1/4}\frac{e^{-(x-x_0)^2/b}}{\sqrt{b}}, 
\label{psi_gw}
&&
	b\equiv 1+\frac{2i\hbar }{m}(t+f_0\sin(\omega t)/\omega)~. 
\label{eq:b}
\end{eqnarray}

\begin{figure}
	\centering
	\includegraphics[scale=1.3]{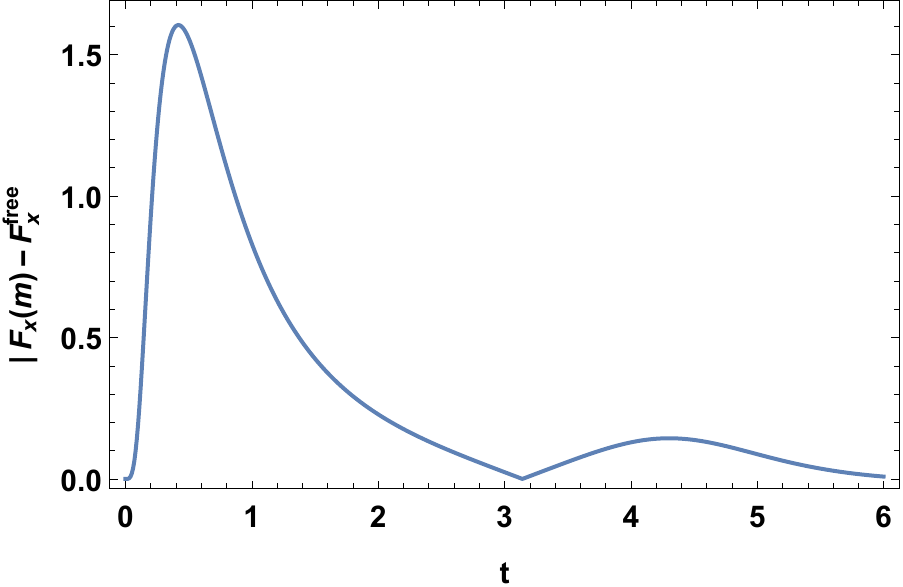}
	\caption{The difference in the mass Fisher information between a free quantum particle and a quantum particle in a GW. Here WEP is violated as  $|F_x (m) - F_x^\text{free}(m)| \neq 0$, and its value fluctuates in time. Units are chosen such that  $\hbar,m,f_0,\omega$ are unity. }
	\label{fig:fisher}
\end{figure}
We can use this wave packet to  calculate the Fisher information in GWs. In Fig.~\ref{fig:fisher}, we  plot the absolute value of the difference between  the mass Fisher information in a GW background $F_x (m)$   and  mass Fisher information for free particle $F_x^{\rm free}(m)$. We observe that even though  $|F_x(m) - F_x^{\rm free}(m)|\neq 0$
(breaking WEP), its value fluctuates over time. So,  we can now define a parameter $\beta^A$ to measure the magnitude  of violation  of WEP for a quantum system  $A$ in GWs  (with Fisher information $F^A_x(m)$) as
\begin{equation}
\beta^A = \frac{\int |F_x^A(m)- F_x^{\rm free}(m)| dt}{\int  |F_x^\text{free}(m)|dt}~. 
\end{equation}
Now for two quantum systems $A$ and $B$,  if $\beta^A > \beta^B$, then the magnitude of violation of WEP is larger  in $A$ than $B$.  

\begin{figure}[H]
  \centering
   \includegraphics[width=.5\linewidth]{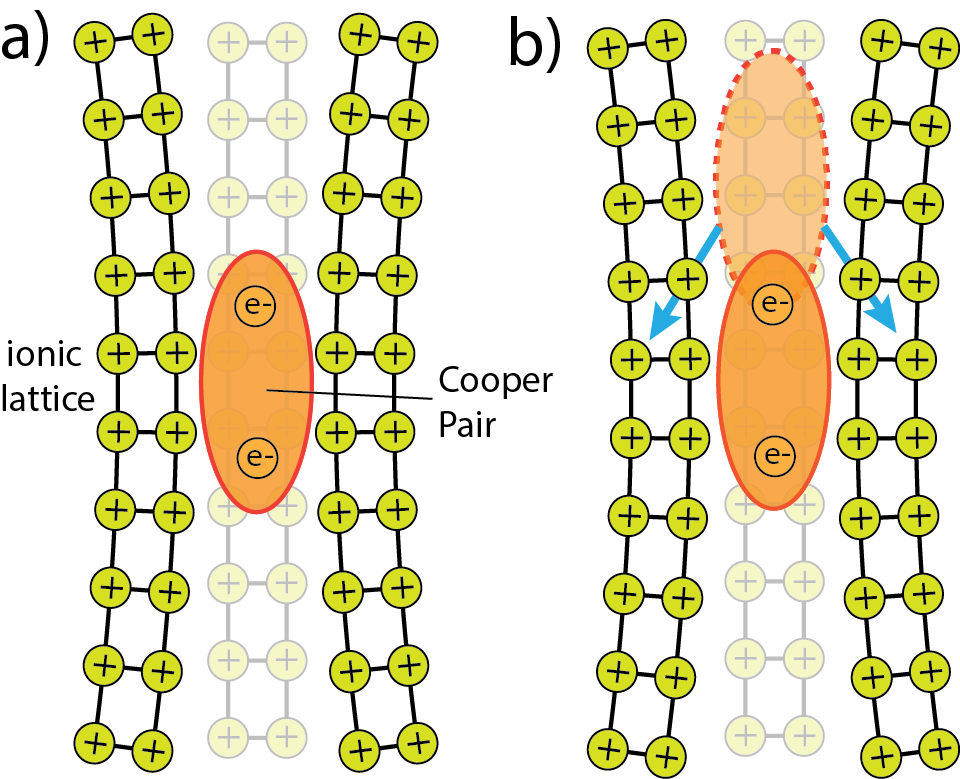}
  \caption{Mechanism for spectral reflection of GWs from superconductors. (a) The negatively-charged CP deforms the positively-charged ionic lattice. (b) The GW accelerates the delocalized CP relative to the lattice. However, the positively-charged ionic lattice suppresses this acceleration, thereby partially reflecting the GW. }
  \label{fig:2}
\end{figure}
This violation of the WEP may be experimentally detected in superconductors because the delocalization of Cooper pairs (CPs) are expected to violate the WEP more than localised particles. This can be observed by  considering the  two-particle wave function of the CP~\cite{tinkham2004introduction}
\begin{equation}
\psi_0(\mathbf{r}^{(1)},\mathbf{r}^{(2)}) = \sum_\mathbf{k}g_\mathbf{k}e^{i\mathbf{k}\cdot\mathbf{r}^{(1)}}e^{-i\mathbf{k}\cdot\mathbf{r}^{(2)}}~,
\label{eq:wavepacket2}
\end{equation}
where $\mathbf{r}^{(i)}$ is the coordinate of particle $i$.
We can now use the unitary transformation operator $U=\exp(-iH_{\rm GW}(\mathbf{p}^{(1)},\mathbf{p}^{(2)})t)$ to get the time-evolution of the CP in a GW background, where
\begin{equation}
 H_{\rm GW}(\mathbf{p}^{(1)},\mathbf{p}^{(2)})=\frac{1}{2m_{\rm cp}}(
 \delta^{ij}+2T^{ij})(p_i^{(1)}p^{(1)}_j+p^{(2)}_ip^{(2)}_j) + m_{\rm cp}c^2 +V~,
\end{equation}
with $V$ as the interaction with the ionic lattice. As the CPs are delocalized relative to the localized ionic lattice  (shown in Fig.~\ref{fig:2}(a)), they   exhibit a larger  magnitude of WEP violation, $\beta^{\rm Cooper} > \beta^{\rm ion}$, because  
\begin{equation}
 \beta^\text{Cooper}=   \frac{\int |F^\text{Cooper}_x(m_{\rm cp}) - F^\text{free}_x(m_{\rm cp})| dt}{\int |F^\text{free}_x(m_{\rm cp})| dt} > 
    \frac{\int  |F^\text{ion}_x(m_{\rm ion}) dt - F^\text{free}_x(m_{\rm ion})| dt}{\int |F^\text{free}_x(m_{\rm ion})| dt} = \beta^{\rm ion}~.
\end{equation}
Thus,  GWs will tend to accelerate the CPs relative to the ionic lattice due to the larger magnitude of  WEP violation in CPs relative to the ionic lattice, as depicted in Fig.~\ref{fig:2}(b). As the CPs and ionic lattice are oppositely charged, they will resist any charge separation, thereby (partially) reflecting the GW, analogous to the way conductors repel electromagnetic waves. It was initially proposed that such a reflection of GWs occurs due to CPs not moving on the geodesics on which ions move  \cite{Minter:2009fx,Inan:2017qdt,Yu:2018fau, Inan:2017ixx}, however, as the system is quantum mechanical, 
the reflection of  GWs in superconductors actually occurs due to the difference in Fisher information for CPs and ions.

Now GWs with wavelengths of the same order as the CP coherence length would be required to test this WEP violation,  but such GWs are difficult to generate in controlled experiments. However, we can  utilize the full spectrum  of the virtual GWs (formed from virtual gravitons in the vacuum) in a gravitational Casimir effect to test  such violations of   WEP  \cite{Quach:2015qwa,Hu:2016lev}. This can be explicitly demonstrated  using the  Einstein field equations linearized around a  flat spacetime metric 
(which resemble the Maxwell equations with a magnetic   monopole)~\cite{matte1953nouvelles,campbell1971debye,Maartens:1997fg,Ramos:2010zza,Szekeres:1971ss}, 
\begin{eqnarray}
\nabla\cdot\mathbf{E}=\kappa\boldsymbol{\rho}^{(E)}~,\label{gem_1}&&
\nabla\cdot\mathbf{B}=\kappa\boldsymbol{\rho}^{(M)}~,\label{gem_2}\nonumber \\
\nabla\times\mathbf{E}=-\frac{\partial\mathbf{B}}{\partial t}-\kappa\mathbf{J}^{(M)}~,\label{gem_3}&&
\nabla\times\mathbf{B}=\frac{\partial\mathbf{E}}{\partial t}+\kappa\mathbf{J}^{(E)}~,\label{gem_4}
\end{eqnarray}
where $\kappa\equiv8\pi G/c^4$,   $E_{ij} \equiv C_{0i0j}$,   $B_{ij}\equiv {\star C_{0i0j}}$,  $C_{\alpha\beta\mu\nu}$ is the Wely tensor, $\star$ denotes Hodge dualization\footnote{Dualization is defined by
  $\star
  C_{\mu\nu\rho\sigma}\equiv\frac12\epsilon_{\mu\nu\alpha\beta}C^{\alpha\beta}_{\rho\sigma}$,  $\star
  J_{\mu\nu\rho}\equiv\frac12\epsilon_{\mu\nu\alpha\beta}
  J^{\alpha\beta}_{\rho}$,
  with  $\epsilon_{\mu\nu\alpha\beta}$ as the Levi-Civita tensor.}, and    ${\rho^{(E)}_i\equiv-J_{i00}}$, ${\rho^{(M)}_i\equiv{-\star J_{i00}}}$, ${J^{(E)}_{ij}\equiv J_{i0j}}$,  ${J^{(M)}_{ij}\equiv\star J_{i0j}}$\footnote{ The    matter current   $J_{\mu\nu\rho}\equiv
      (\eta_{\rho[\mu}T_{,\nu]}/3) -T_{\rho[\mu,\nu]}$ is obtained from stress perturbations   $T_{\mu\nu}$ ($T\equiv T^\mu_\mu$).}. 
  Now for parallel superconducting plates separated by a distance $b$ along the $z$-axis (with  ${k}_{\parallel}\equiv\sqrt{k_x^2 +k_y^2}$), the gravitational Casimir energy  can be obtained by summing over  ($\omega_n^+$, $\omega_n^\times$), which are the relevant   modes of virtual GWs for the system   ~\cite{Bordag:2009zzd, Hu:2016lev},  
\begin{equation}
E_0(b)=\frac{\hbar}{4\pi}\int_0^\infty k_{\parallel} dk_{\parallel} \sum_{n}(\omega_n^++\omega_n^\times)\sigma~\label{E0}~,
\end{equation} 
where $\sigma$ is the surface area of the plates.
It may be noted that it diverges due to the  summation over the infinite number of allowed
modes, and we need to renormalize it  by subtracting the Casimir  energy at infinite separation 
\cite{Quach:2015qwa}
\begin{equation}
E_{R}(b) =\frac{E_0}{\sigma}-
\lim_{b \to\infty}\frac{E_0}{\sigma}~.\\
\end{equation}
It is this finite renormalized  gravitational Casimir energy that can be detected using superconducting parallel plates \cite{norte2018platform}. The strength of the gravitational Casimir force will depend on the difference between the  magnitude  of WEP violation experienced by the delocalized CP ($\beta^{\rm Copper}$) and the localized ionic lattice ($\beta^{\rm ion}$).  Although the  gravitational Casimir effect  is expected to be smaller than the electromagnetic Casimir effect, it can be measured from the additional force produced at the onset of superconductivity (which is experimentally easier than measuring absolute Casimir forces). In fact,   the magnitude of this gravitational Casimir effect has been estimated to be on the same order as the electromagnetic Casimir effect using heuristic calculations  \cite{Minter:2009fx,Inan:2017qdt,Yu:2018fau, Inan:2017ixx, Quach:2015qwa,Hu:2016lev}. However, as it is important to determine the exact magnitude of this gravitational Casimir effect,   we need to rigorously perform these calculations. It may be noted that  even if the exact magnitude of the gravitational Casimir effect  is reduced by several orders of magnitude, from what has been calculated using heuristic calculations, it would still be measurable.

\begin{figure}[H]
  \centering
    \includegraphics[width=1\linewidth]{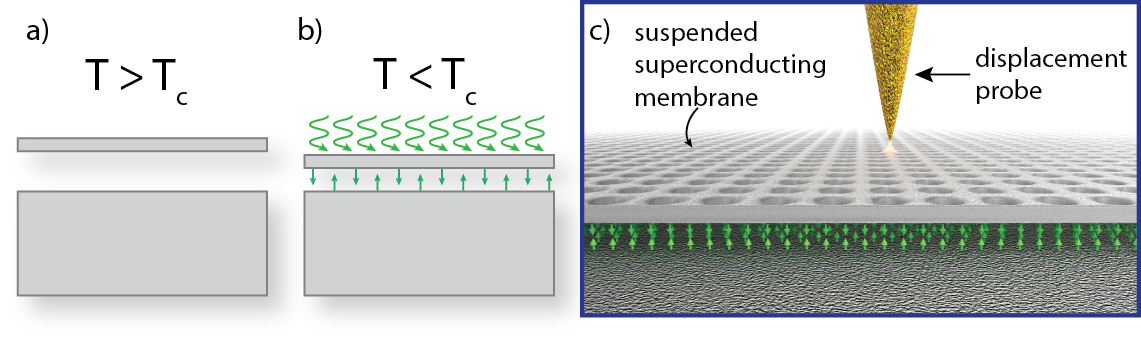}
  \caption{General schematic of experiments: a) A superconducting nanomembrane is fabricated above a superconducting substrate. b) Once cooled below the superconducting transition temperature $T_c$, one would observe the sudden change in distance between membrane and substrate due to the gravitational Casimir effect. 
  c) One could measure this change in displacement using a number of currently available displacement probes, which can measure distances on the atomic scale.
 }
  \label{fig:3}
\end{figure}

A gravitational Casimir experiment with superconductors would be relatively simple, as shown in Fig~\ref{fig:3}. On a microchip, one can achieve remarkable parallelism~\cite{norte2018platform}, making it possible to fabricate a free-standing superconducting nanomembrane over a superconducting substrate \cite{Moura18}. The entire microchip can then be cooled below the superconducting transition temperature $T_c$, experiencing an additional gravitational Casimir  
force, 
as depicted in Fig~\ref{fig:3}(a, b).  
Any small additional forces at $T_c$ would lead to large displacements of these ultra-sensitive nanomembranes due to their high-aspect-ratio \cite{Moura18}. 
A displacement probe such as those used in scanning tunneling microscopy \cite{battisti2018}, or a photonic crystal probe \cite{magrini2018near}, could measure displacements of this membrane with atomic precision, as depicted in Fig~\ref{fig:3}(c). The magnitude of WEP violation in CPs depends on the CP coherence length, which in turn depends on the properties of  superconductors~\cite{lee2014interfacial,bednorz1986possible}, so one may use various superconducting materials to measure the gravitational  Casimir effect with current technology. As the WEP is a building block of general relativity, convincingly demonstrating its violation would signify a drastic departure from our current understanding of gravity at short distances.

\section*{Acknowledgements}
SB is supported by Mobilitas Pluss N$^\circ$ MOBJD423 by the Estonian government. JQQ acknowledges the Ramsay fellowship for financial support of this work. RAN acknowledges the NWO Start-Up Fellowship for support of this work and Sander Otte and Simon Groeblacher for fruitful discussion. The rendering for Fig 3c is courtesy of Moritz Forsch.

\bibliographystyle{utphys}
\bibliography{references}

\end{document}